\documentclass[12pt]{article} 
\pdfoutput=1
\usepackage{amssymb,graphicx}
\usepackage{epstopdf}
\usepackage{amsmath,amsfonts}
\usepackage{epsfig} 
\usepackage{graphicx,graphics}

\begin{document}

\title{\bf Statistical anisotropy of CMB as a
probe of conformal rolling scenario}
\author{S.~R.~Ramazanov$^{a,b}$\footnote{{\bf e-mail}: sabir\_ra@nbi.dk} ,
G.~I.~Rubtsov$^{b,c}$\footnote{{\bf e-mail}: grisha@ms2.inr.ac.ru}
\\
$^a$ \small{\em Niels Bohr International Academy and Discovery Center, Niels Bohr Institute,}\\
\small{\em University of Copenhagen, Blegdamsvej 17, DK-2100, Copenhagen, Denmark}\\
$^b$ \small{\em Physics Department, Moscow State University,}\\
\small{\em Vorobjevy Gory, 119991, Moscow, Russia}\\
$^c$ \small{\em Institute for Nuclear Research of the Russian Academy of Sciences} \\
\small{\em Prospect of the 60th Anniversary of October 7a, Moscow, Russia, 117312}\\
} 
\maketitle

\begin{abstract} 
Search for the statistical anisotropy in the CMB data is a powerful tool 
for constraining models of the early Universe. In this paper we focus on the recently proposed 
cosmological scenario with conformal rolling. We consider two sub-scenarios, one of which involves a long intermediate 
stage between conformal rolling and conventional hot epoch. Primordial 
scalar perturbations generated within these sub-scenarios have different
direction-dependent 
power spectra, both characterized by a single parameter $h^2$. 
We search for the signatures of this anisotropy
in the seven-year WMAP data using quadratic maximum likelihood method, first applied for similar
purposes by Hanson and Lewis. 
We confirm the large quadrupole anisotropy detected in $V$ and $W$ bands, 
which has been argued to originate from systematic effects rather than from cosmology. We construct an estimator 
for the parameter $h^2$. In the case of the sub-scenario with the intermediate stage we set an
upper limit $h^2 < 0.045$ at the 95\% confidence level. The constraint on $h^2$ is much weaker 
in the case of another sub-scenario, where the intermediate stage is absent. 
\end{abstract}

\section{Introduction and summary} 

Recent advances in the observational cosmology make it real to start probing
the most intriguing aspects of the 
Universe. In particular, it is of importance to  inquire whether the statistical 
isotropy of the scalar perturbations is exact or only approximate. This issue 
is of special interest because the statistical isotropy is one of the key assumptions of 
the six-parametric $\Lambda$CDM model
and is favored by inflation. Thus, the violation 
of this property in the observed CMB 
would imply a highly non-trivial extension of the 
now standard cosmological model. 
An additional motivation to search for the statistical anisotropy is the possible presence
of various anomalies in the CMB data, such as 
alignment of low multipoles, axis of evil, power assymetries, cold spots and others~\cite{Tegmark}--\cite{Rath}.

In the statistically anisotropic but spatially homogeneous Universe, 
the power spectrum of the primordial scalar perturbations $\zeta ({\bf k})$ depends 
on the direction of the wave vector ${\bf k}$. The power spectrum can then be written as follows, 
\begin{equation} \label{power}
{\cal P}_{\zeta} ({\bf k})={\cal P}_0 (k) \left[ 1+a(k) \sum_{LM} q_{LM} Y_{LM} (\hat{{\bf k}})\right] \; ,
\end{equation}
where $\hat{\bf k}= {\bf k}/{k}$. 
The coefficients $q_{LM}$ parametrize the direction-dependent part, which one expands in spherical harmonics
 $Y_{LM} (\hat{{\bf k}})$. Unlike in Ref.~\cite{Pullen}, we assume here
that the dependence on the wavenumber $k$ may be absorbed into one function $a(k)$. Commutativity
of the classical field $\zeta({\bf x})$ yields ${\cal P}_{\zeta} ({\bf k}) = {\cal P}_{\zeta} (-{\bf k})$
and hence $q_{LM} = 0$ for odd $L$.

The generic prediction of the inflationary theory is that the power spectrum is isotropic,
$a(k)=0$.
However, the statistical anisotropy can be generated in models of inflation involving 
vector fields~\cite{Ackerman, Soda, Dimopoulos, Wagstaff} 
or scalar fields with non-minimal kinetic terms~\cite{Armendariz}; for reviews 
see also Refs.~\cite{Konstantinos} and~\cite{Soda2}. Somewhat more exotic examples are given by introducing 
Bianchi I geometry~\cite{Peloso} or noncommutative field theory~\cite{Akofor, Koivisto}.
The most common feature of these models is the statistical anisotropy of a special quadrupole type 
(the only non-vanishing coefficient in a certain reference frame on the
celestial sphere is $q_{20}$). 
This prediction arises, e.g.,
in the model with the rotational invariance broken by a space-like vector~\cite{Ackerman}, or in the hybrid inflation 
incorporating a vector field coupled to 
the waterfall scalar~\cite{Soda}. 
However, higher multipoles $q_{LM}$ can also emerge within the inflationary framework, see, e.g., Ref.~\cite{Armendariz}.

In inflationary theory, the observed approximate flatness of the scalar power spectrum is due to
the approximate de~Sitter symmetry of inflating space-time. It has been suggested some time ago~\cite{Antoniadis}
that, alternatively, the flatness of the power spectrum may be a consequence of conformal symmetry.
Concrete models of this sort have been proposed in Refs.~\cite{Rubakov,Trincherini} and 
further developed in Refs.~\cite{Osipov}--\cite{Hinterbichler}. It is this class of models that we focus on in this paper.

In models of 
Refs.~\cite{Rubakov,Trincherini} and similar ones, it is assumed that the cosmological evolution
starts from or passes through a conformally invariant state with effectively flat geometry.
This state is unstable, and conformal symmetry $SO(4,2)$ gets broken down to $SO(4,1)$ by a time-dependent
(rolling) scalar field. During this conformal rolling stage, another field of zero conformal weight
develops perturbations which automatically have flat power spectrum. These perturbations are reprocessed
into adiabatic perturbations by one or another mechanism (e.g., of Refs.~\cite{Linde:1996gt,Dvali:2003em})
at some later epoch, after the end of conformal rolling. It is then natural that the adiabatic perturbations
inherit the properties of the original field perturbations, modulo possible additional non-Gaussianity.

At the conformal rolling stage, the properties of perturbations to both linear and leading non-linear orders
are uniquely determined by the underlying conformal symmetry~\cite{Hinterbichler}, modulo the overall
amplitude and a single dimensionless parameter which we call $h^2$ (in the model of Ref.~\cite{Rubakov},
the amplitude is also determined by $h^2$, see Section~\ref{sec:fields}). 
This parameter governs the non-Gaussianity and statistical anisotropy. 
The statistical anisotropy generated in the conformal rolling scenario is quite
different from the predictions of inflation. 
In particular, the coefficients 
$q_{LM}$ parametrizing the power spectrum~\eqref{power} are 
the random variables rather than fixed parameters 
of the model. 

There are two sub-scenarios of the
conformal rolling scenario which differ by the behaviour of the cosmologically interesting 
modes after the end of the
conformal rolling stage. One possibility is that these modes are already superhorizon
by the end of conformal rolling and remain frozen until the late hot epoch. The second sub-scenario assumes
that there is an intermediate stage at which the field perturbations evolve 
in a non-trivial way before crossing out the
cosmological horizon and getting finally frozen out. 
In the latter case, the evolution of the field 
perturbations results in the statistical anisotropy of all 
even multipoles. Notably, it does not depend on the magnitude of the momentum ${\bf k}$, i.e.,
$a(k)$ is independent of $k$ \cite{Ramazanov}. 
The predictions of the former sub-scenario for the
statistical anisotropy are considerably different. To the linear order 
in the parameter $h$, one obtains the statistical anisotropy of the general quadrupole type. It is characterized 
by the amplitude which decreases with the wavenumber $k$, i.e. $a(k) \propto k^{-1}$~\cite{Libanov}. This means that the 
corresponding effects in the CMB sky are suppressed at high CMB multipoles $l$. 
The statistical anisotropy of the 
special quadrupole type also appears in the next-to-linear order. Though the effect is suppressed by the additional 
power of the parameter $h$, now the amplitude $a(k)$ does not depend on the wavenumber $k$~\cite{Libanov}. Hence, 
the subleading contribution may make stronger imprint on the CMB.

Signatures of the statistical anisotropy in the CMB temperature maps have been searched for
in Refs.~\cite{Groeneboom}--\cite{Eriksen}.
Motivated by the model of Ref.~\cite{Ackerman}, Groeneboom and Eriksen~\cite{Groeneboom} 
discovered the evidence for the quadrupole statistical anisotropy in the 
five-year WMAP data. However, it was found to be 
nearly aligned with 
the ecliptic poles. Using quadratic maximum likelihood estimator, Hanson and Lewis~\cite{Duncan} extended the analysis 
to higher multipoles. They also included the relevant prefactor in the covariance neglected 
in Refs.~\cite{Ackerman} and~\cite{Groeneboom}. Hanson and Lewis~\cite{Duncan}
confirmed the result on the large quadrupole $q_{2M}$ lying nearly in the ecliptic plane. 
The strongest indication of the statistical isotropy violation, non-zero at the $9\sigma$ confidence level, was 
found in the $W$ band of the five-year WMAP data in Ref.~\cite{Eriksen}.
These findings have been confirmed by the WMAP team~\cite{WMAP-7an} in their analysis
of the seven-year data.
One possible explanation of the anomalous quadrupole is the systematics 
inherent in the WMAP data. 
As argued in Ref.~\cite{Challinor}, 
large observed statistical anisotropy may result from beam asymmetries 
rather than have the cosmological origin.

The purpose of this paper is to constrain the parameter $h^2$ of the conformal rolling scenario 
from the non-observation of the cosmological statistical anisotropy in the seven-year WMAP data. 
We follow the general method proposed by Hirata and Seljak 
for the purpose of studying 
CMB lensing 
and known as the quadratic maximum likelihood (QML) estimation~\cite{Hirata}. 
As discussed in Ref.~\cite{Duncan}, the same idea can be applied to the study of the statistically 
anisotropic properties of CMB. 
In this case one assumes that the coefficients $q_{LM}$ are small and expands the log-likelihood of the observed CMB 
to the 
second order in these parameters. By maximizing the log-likelihood with respect to the coefficients $q_{LM}$, 
one obtains the estimator. Results derived within the QML approximation are in a good agreement with the exact 
likelihood methods. 

We apply this method to construct the estimator for the parameter $h^2$. In view of the results quoted,
the estimated values are expected to be inconsistent 
with the statistical isotropy because of the alleged systematics present in the WMAP data. 
Assuming that the interpretation in terms of systematics is correct, we set the upper limits 
on the parameter $h^2$ in the following way. For each value of $h^2$, we simulate the parameter sets $\{ q_{LM} \}$, and then 
generate a number of anisotropic maps for each set $\{ q_{LM} \}$. 
From the maps generated, we estimate the values of the parameter $h^2$. We require that in 95\% cases
they should not exceed 
the value estimated from the observed CMB. In this way we constrain the conformal rolling
sub-scenario with the intermediate stage,
\begin{equation}
\nonumber 
h^2 < 0.045 
\end{equation}
at the 95\% confidence level. 
The constraint is much weaker in the framework of the alternative sub-scenario. 
The reason is that the amplitude of the leading order quadrupole decreases as $k^{-1}$. This translates 
into the suppression of the statistical anisotropy effects at high CMB multipoles. Thus, 
the data useful for the analysis are effectively limited,
statistical errors are large and the 
constraint is 
$h^2 < 190 $
at the 95\% confidence level. The constraint is improved significantly, 
once we take into account 
the subleading contribution to the statistical anisotropy. This contribution is 
of the special quadrupole type, and has the amplitude $a(k)$, which is independent 
of the wavenumber $k$. Thus, the number of CMB multipoles useful in the analysis is 
much larger. This somewhat compensates the smallness 
of the constant $h$, and we obtain the stronger constraint,
\begin{equation}
\nonumber 
h^2 \ln \frac{H_0} {\Lambda} <7 \;  
\end{equation}
at the 95\% confidence level. Here $H_0$ is the present value of the Hubble parameter, which plays the role 
of the ultraviolet cutoff, and $\Lambda$ is the infrared 
cutoff. Without going into speculation on the value of the constant $\Lambda$, we 
point out that this constraint 
 is still very weak, in view of the fact that  
the conformal rolling scenario is self-consistent only at $h^2 \ll 1$ anyway.

We conclude that the statistical anisotropy is the relevant signature 
of the conformal rolling with the intermediate stage. It is of particular interest in 
view of the upcoming Planck data. Hopefully, the latter will be free 
of the quadrupole anomaly. The other expected advantage of the Planck data 
is the larger range of the CMB multipoles, which translates into smaller statistical
errors. These two 
factors are expected to improve the sensitivity of the data to the parameter $h^2$ by more than 
an order of magnitude. On the other hand, statistical 
anisotropy appears to be a weak signature of the alternative sub-scenario, and the Planck data 
are not expected to improve the situation significantly. Thus, it makes sense 
to focus on the other prediction of this sub-scenario, the non-Gaussianity~\cite{Mironov2, Mironov1}. 
At the level of bispectrum, the non-Gaussianities in the
conformal scenario are not particularly special. The shape of the
intrinsic bispectrum is dictated~\cite{Hinterbichler:2012mv} by the symmetry breaking pattern 
$SO(4,2) \to SO(4,1)$ and coincides with the bispectrum of a spectator massless
scalar field in inflationary theory~\cite{Zaldarriaga:2003my,Creminelli:2011mw}
(in fact, the intrinsic bispectrum may vanish for symmetry reasons, see Section~\ref{sec:fields}). 
The non-Gaussianity 
generated at the conversion epoch is not specific to the conformal scenario either.
So, bispectrum alone cannot discriminate between the conformal scenario and, say,
inflation equipped with the curvaton mechanism. 
On the other hand,
the non-Gaussianity of a rather peculiar form arises in the trispectrum. 
Existing constraints (see, e.g. Ref.~\cite{Fergusson}) are model-dependent and cannot be directly applied  
to our model. We leave for the future the analysis of the 
CMB data aiming at the search for the non-Gaussianity characteristic of the conformal scenario.


This paper is organized as follows. In Section~\ref{sec:fields} we review the properties of 
the field perturbations at the conformal rolling stage. The power spectrum of relevant perturbations 
generated by the end of their evolution posseses directional dependence. We show this explicitly 
in Section~\ref{statan}. We consider two sub-scenarios: one in which the cosmologically interesting modes 
are superhorizon 
by the end of the conformal rolling 
stage (Section~\ref{sec:scenarioA}) and another, with  long intermediate stage 
(Section~\ref{sec:scenarioB}). In Section~\ref{sec:estimator} we review the 
main ideas of the QML method and construct  model-independent estimators for the coefficients $q_{LM}$.
We also construct an estimator for the parameter $h^2$. Section~\ref{sec:results} contains our main results.
We
implement the estimators to the seven-year WMAP data and constrain the parameter $h^2$ of the two sub-scenarios.

\section{Conformal rolling scenario} 
\subsection{Fields and their perturbations at conformal rolling}
\label{sec:fields}

The main ingredient of the conformal rolling scenario~\cite{Rubakov,Trincherini} is the conformal 
stage preceding the conventional hot epoch. As we point out below, the properties of the
conformal rolling stage are quite general~\cite{Mironov1,Hinterbichler}, as they are unambiguously determined
by conformal symmetry. 
Yet it is instructive to review a simple explicit model~\cite{Rubakov} illustrating this scenario.
It involves
a complex scalar field $\phi$, conformally coupled 
to gravity, which rolls down the negative quartic potential $V(\phi)=-h^2 |\phi|^4$, where
$h$ is a small parameter. 
At large field values, the potential  
is assumed to change and have a minimum at $|\phi|=f_0$ where conformal symmetry is explicitly broken. 
The field $\phi$ is a spectator at the conformal rolling stage and somewhat later,
i.e., the evolution of the Universe is dominated by some other matter (see Ref.~\cite{Hinterbichler}
for an alternative version where $\phi$ is the dominant matter component).

The central object in the model is the phase $\theta$ defined by $\phi = |\phi| \mbox{e}^{i\theta}$.
Its
perturbations $\delta \theta$ 
start off 
as vacuum fluctuations and eventually freeze out. By the end of conformal rolling, perturbations $\delta \theta$ have 
flat power spectrum. Once the radial 
field $|\phi|$ settles down to $f_0$ and conformal symmetry gets broken, what remains are the perturbations of the phase, 
which at this point are isocurvature 
perturbations. They get reprocessed into adiabatic perturbations at much later epoch. 
In this way the flat power spectrum of the adiabatic perturbations is
obtained. It can be slightly tilted if there is small explicit breaking of conformal symmetry
at the conformal rolling stage~\cite{Osipov}. 

Let us discuss conformal rolling in more detail.
At this stage, the theory is described by the action 
\begin{equation}
\nonumber
S=S_{G+M}+S_{\phi} \; ,
\end{equation}
where $S_{G+M}$ is the action for gravity and some matter that dominates the evolution of the 
Universe, and 
\begin{equation}
\label{feb15-2}
S_{\phi}=\int d^4 x \sqrt{-g} \left[g^{\mu \nu} \partial_{\mu} \phi^{\star} \partial_{\nu} \phi +
\frac{R}{6} \phi^{\star} \phi -V(\phi)\right]
\end{equation}
is the action for the complex scalar field. Assuming that the background metric is homogeneous, 
isotropic and spatially flat, 
one introduces the field $\chi =a\phi$, where $a$ is the scale factor, 
and obtains its action in conformal coordinates in the Minkowskian form, 
\begin{equation}
\nonumber
S[\chi] =\int d^3 x d\eta \left(\eta^{\mu \nu}\partial_{\mu} \chi^{\star} \partial_{\nu} \chi +h^2 |\chi|^4\right) \; .
\end{equation} 
Here $\eta^{\mu \nu}$ is the Minkowski metric and $\eta$ is the conformal time.

Since the scalar potential is negative, the conformally invariant state $\chi=0$ is unstable,
and the field rolls down its potential. 
One assumes that the background $\chi_c$ is spatially homogeneous and 
without loss of generality chooses the background solution real. It is given by 
\begin{equation}
\nonumber
\chi_{c} (\eta) =\frac{1}{h(\eta_{\star} -\eta)} \; ,
\end{equation}
where $\eta_{\star}$ is a free parameter, which is interpreted as the end-of-roll time. 
Let us consider perturbations in this background. 
To the leading order in $h$, perturbations 
$\delta \chi_{1} =\sqrt{2} \delta \mbox{Re} \chi$ and $\delta \chi_2 =\sqrt{2} \mbox{Im} \chi$ decouple from each other. 
We begin with the radial perturbations $\delta \chi_1$. 
They obey the linearized field equation, in momentum representation, 
\begin{equation}
\nonumber
(\delta \chi_1)^{''} +p^2 \delta \chi_1 -6h^2 \chi^2_c \delta \chi_1=0 \; .
\end{equation}
The properly normalized solution is 
\begin{equation}
\nonumber
\delta \chi_1 =\frac{1}{4\pi} \sqrt{\frac{\eta_{\star} -\eta}{2}}H^{(1)}_{5/2} [p(\eta_{\star} -\eta)] B_{{\bf p}}+h.c.\; , 
\end{equation}
where $B_{\bf p}$ and $B^{\dagger}_{\bf p}$ are annihilation and creation operators obeying the canonical 
commutational relations;
$H^{(1)}_{5/2}$ is the Hankel function. At late times the solution approaches the asymptotics 
\begin{equation}\label{realperturb}
\delta \chi_{1} =\frac{3}{4\pi^{3/2}} \frac{1}{p^{5/2} (\eta_{\star} -\eta)^2} B_{\bf p} +h.c. \; .
\end{equation}
The behaviour $\delta \chi_1 \sim (\eta_{\star} -\eta)^{-2}$ is interpreted as a shift of the end-of-roll 
time $\eta_{\star}$, which now becomes a random field. Indeed, with perturbations included, the radial field
$\mbox{Re} \chi=\chi_{c} +{\delta \chi_1}/{\sqrt{2}}$ can be written at late times as follows 
\begin{equation}
\label{feb15-1}
\mbox{Re} \chi =\frac{1}{h[\eta_{\star} ({\bf x})-\eta]} \; ,
\end{equation}
where 
\begin{subequations}
\begin{align}
\eta_{\star} ({\bf x}) &=\eta_{\star} + \delta \eta_{\star} ({\bf x}) \; ,
\\
\label{feb15-3}
\delta \eta_{\star} ({\bf p}) &=- \frac{3h}{4\sqrt{2}\pi^{3/2} p^{5/2}}  B_{\bf p} +h.c. \; .
\end{align}
\end{subequations} 
As it stands, the expression (\ref{feb15-1}) involves all powers of the small
parameter $h$, i.e., it appears
to imply some resummation. However, we will be primarily interested in the first
non-trivial order in $h$, so we understand Eq.~(\ref{feb15-1}) merely as a convenient
book-keeping tool: to the first non-trivial order in $h$  the right
hand side of Eq.~(\ref{feb15-1}) is equivalent to $\chi_c + \delta \chi_1$.
It is natural to assume that the fields are initially in their vacuum state.
Then 
the field $\delta \eta_{\star} ({\bf x})$ is a 
time-independent Gaussian field with red power spectrum. Clearly, 
the overall spatially homogeneous 
shift of the end-of-roll time is irrelevant, since it can be absorbed 
into the redefinition of $\eta_{\star}$. What is important 
is the gradient of $\eta_{\star} ({\bf x})$, 
\begin{equation}
\label{feb15-30} 
v_i=-\partial_{i} \eta_{\star} ({\bf x})\; .
\end{equation}
It has flat power spectrum, while the higher derivatives have blue spectra.

Let us turn to the object of primary interest, namely, the perturbations of the phase $\theta$ or,
equivalently, 
perturbations $\delta \chi_2$ of the imaginary part of the field. 
We are interested in the leading and subleading orders in the small parameter
$h$, so we take the (real) background in the form (\ref{feb15-1}).
The linearized field equation for $\delta \chi_2$ reads 
\begin{equation} \label{chi2eq}
(\delta \chi_2)^{''}-\partial_i \partial_i \delta \chi_2 -\frac{2}{[\eta_{\star} ({\bf x}) -\eta]^2} \delta \chi_2 =0\; .
\end{equation}
At early times, when $k(\eta_{\star}-\eta) \gg 1$, we get back to the Minkowskian massless equation. 
The solution to Eq.~\eqref{chi2eq} has the following form, 
\begin{equation}
\nonumber
\delta \chi_2 ({\bf x}, \eta) =\int \frac{d^3 k}{(2\pi)^{3/2} \sqrt{2k}} (\delta \chi^{(-)}_2 ({\bf k}, {\bf x}, \eta) A_{\bf k} +h.c.)\; ,
\end{equation}
where $\delta \chi^{(-)}_2 ({\bf k}, {\bf x}, \eta)$ tends to $e^{i{\bf kx}-i k\eta}$ as $\eta \rightarrow - \infty $ 
and $A_{\bf k}$, 
$A^{\dagger}_{\bf k}$ is another set of annihilation and creation operators. Modulo corrections proportional to 
$\partial_{i} \partial_{j} \eta_{\star}$ and $v^2$, the solution with this initial condition 
is~\cite{Libanov} 
\begin{equation}
\nonumber
\delta \chi^{(-)}_2 ({\bf k}, {\bf x}, \eta)= -e^{i{\bf kx}-ik\eta_{\star} ({\bf x})-i{\bf kv} (\eta_{\star}-\eta)} 
\sqrt{\frac{\pi}{2} q (\eta_{\star} ({\bf x})-\eta)}H^{(1)}_{3/2}[q ({\eta_{\star} ({\bf x}) -\eta)]} \; , 
\end{equation}
where  $q=k+{\bf kv}$ and ${\bf v}$ is given by Eq.~\eqref{feb15-30}. 
At late times one has $\delta \chi_2 \sim [\eta_{\star} ({\bf x}) -\eta]^{-1}$. As a result, 
the phase 
perturbations 
$\delta \theta = \delta \chi_2/(\sqrt{2} \chi_1)$
freeze out. In this regime,
the phase perturbations, including 
corrections of  order $\partial_i \partial_j \eta_{\star}$ and $v^2$,
 have the following form~\cite{Libanov}:  
\begin{equation} \label{phaseperturb}
\delta \theta ({\bf x}, \eta)=\int \frac{d^3 k}{\sqrt{k}} 
\frac{h}{4\pi^{3/2} k} e^{i\gamma^{-1} k_{||} x_{||}+i{\bf k}^{T} {\bf x}^{T}} A_{\bf k} \left(1-\frac{\pi}{2k} \frac{k_i k_j}{k^2} 
\partial_i \partial_j \eta_{\star} +\frac{\pi}{6k} \partial_{i} \partial_{i} \eta_{\star} \right) \; ,
\end{equation}
where $\gamma=(1-v^2)^{-1/2}$ is the standard Lorentz factor; $k_{||}$ and ${\bf k}^{T}$ denote the 
momenta in the direction of the ``velocity'' ${\bf v}$ and in the orthogonal direction, respectively.
These are the phase perturbations by the end of conformal rolling. As promised, 
they do not depend on the conformal time $\eta$, 
\begin{equation}\label{freezeout}
\partial_{\eta} \delta \theta ({\bf x}, \eta) |_{\eta =\eta_{\star} ({\bf x})}=0 \; .
\end{equation}
To the leading order in $h$, the parameter $\eta_*$ is independent of ${\bf x}$, so that
${\bf v}=0$, $\partial_i \partial_j \eta_* = 0$ and the phase perturbations \eqref{phaseperturb} are Gaussian random field
with the flat power spectrum and amplitude of order $h$. To the subleading order, the expression 
\eqref{phaseperturb} involves another time-independent Gaussian field $\delta \eta_*$. This leads to
both non-Gaussianity and statistical anisotropy of $\delta \theta ({\bf x})$ and hence
$\zeta ({\bf x})$. Note that the symmetry $\theta \to -\theta$ ensures that the intrinsic bispectrum vanishes
in the concrete model we review.

An important remark is in order. Even though we illustrated the conformal rolling mechanism
by making use of the concrete model \eqref{feb15-2}, the results are characteristic of
the entire class of conformal models. As an example, the above formulas are valid~\cite{Mironov1},
modulo field redefinition, in the Galilean Genesis model~\cite{Trincherini} based on conformal Galileon 
field with higher derivative action~\cite{Nicolis:2008in}. In fact, these formulas hold~\cite{Hinterbichler},
provided that the theory has the following general properties at the conformal rolling stage: (i) space-time is
effectively Minkowskian; (ii) conformal symmetry $SO(4,2)$ is spontaneously broken down to
$SO(4,1)$ by a homogeneously rolling scalar field of non-zero conformal weight ($\chi_1$ in the above example);
(iii) there is another scalar field of zero conformal weight (the phase field $\theta$ above) whose
perturbations are in the end converted into the adiabatic perturbations. The only qualification is that
the latter field need not be compact, so the amplitude of its canonically normalized
perturbations can be arbitrary, whereas the overall amplitude of $\delta \theta$ in the
model \eqref{feb15-2} is
proportional to the same parameter $h$ that determines the amplitude of the perturbations
$\delta \eta_*$ and hence non-linear terms in $\delta \theta$, see Eqs.~\eqref{feb15-3} and \eqref{phaseperturb}. 

To avoid confusion, we note that the model (\ref{feb15-2}) is different from the
Galilean Genesis model~\cite{Trincherini} and the pseudo-conformal model of 
Ref.~\cite{Hinterbichler} in that in the latter models, the cosmological background
itself is driven by the rolling scalar field ($\chi_1$ in our notations).
Therefore, the perturbation of this field (mode $\delta \chi_1$) is gauge-dependent,
while the cuvature perturbation $\zeta$ has blue power spectrum. This, however,
does not invalidate the discussion in the previous paragraph. Indeed, the
time shift (analogous to our $\delta \eta_*$) can be given a gauge-invariant
definition~\cite{Trincherini}; this field is related to, but different from $\zeta$. 
Furthermore, at early times, when the conformal mechanism operates in models
of Refs.~\cite{Trincherini,Hinterbichler}, the energy density and pressure
of the rolling field
are small, so the effects due to gravity are negligible. This is particularly clear in
the Newtonian gauge~\cite{Trincherini,LibanovRubakov}, where the metric perturbations are
small at early times, while the perturbations in the rolling field have the form equivalent
to Eq.~\eqref{realperturb}. In this regime, the curvature perturbation $\zeta$ is
irrelevant, as long as one is interested in the perturbations of the field similar
to our phase $\theta$, and the formulas of this Section remain valid. 
This has been shown expicitly~\cite{LibanovRubakov} 
in the pseudo-conformal
model of Ref.~\cite{Hinterbichler}, both in the Newtonian gauge and in the gauge
$\delta \chi_1 = 0$.

\subsection{Two sub-scenarios and statistical anisotropy of primordial perturbations} \label{statan}

We continue to use the terminology borrowed from the model \eqref{feb15-2}.

At the conformal rolling stage, the phase perturbations $\delta \theta$ get frozen out
because of the fast evolving background \eqref{feb15-1}. Once the conformal rolling stage ends and
the radial field settles down to a constant value $f_0$, the phase becomes a scalar field
minimally coupled to gravity. The behavior of its perturbations at later times depends on whether they are
super- or subhorizon in the conventional sense by the end of conformal rolling. Accordingly, there are two
sub-scenarios which give different predictions for the statistical anisotropy (and non-Gaussinaity 
as well)~\cite{Mironov2, Mironov1, Ramazanov}.

\subsubsection{Sub-scenario A}
\label{sec:scenarioA}

In this sub-scenario, the phase perturbations are already superhorizon  by the end of conformal rolling.
So, they remain frozen out, and there is a simple relation between the adiabatic and phase perturbations,
$\zeta ({\bf x}) =\mbox{const} \cdot \delta \theta ({\bf x})$ plus possible non-linear terms, where
$\delta \theta ({\bf x})$ are the phase perturbations late at the conformal rolling stage. So, the properties of $\zeta$
can be read off from  Eq.~~\eqref{phaseperturb}, modulo possible non-Gaussianity generated when $\delta \theta$
is converted into $\zeta$.

The interaction of the phase perturbations with the radial ones at the conformal rolling stage 
leads to non-trivial effects in the spectrum of the primordial perturbations. 
In particular, it gives rise to the statistical anisotropy. Indeed, let us 
consider the two-point product $\delta \theta ({\bf x}) \delta \theta ({\bf x}')$ and average 
it over the realizations of the operators $A_{\bf k}$ and $A^{\dagger}_{\bf k}$. To the 
leading order, we obtain the 
flat and isotropic power spectrum. The directional dependence appears once we take into account 
corrections coming from the derivatives of the end-of-roll time $\eta_* ({\bf x})$
and keep only those modes of $\delta \eta_* ({\bf x})$ which are still superhorizon today
(shorter modes of $\delta \eta_* ({\bf x})$ give rise to the non-Gaussianity rather than
statistical anisotropy~\cite{Mironov2,Mironov1}). For so long modes of $\delta \eta_* ({\bf x})$, 
it does not make sense to average over the realizations of
the operators $B_{\bf p}$, $B^\dagger_{\bf p}$ at this stage. 
In this way one obtains the power spectrum of the 
primordial perturbations $\zeta ({\bf k})$~~\cite{Libanov}: 
\begin{equation}
\label{feb15-20}
{\cal P}_{\zeta} ({\bf k})={\cal P}_0 (k)\left(1+ Q_{1} ({\bf k}) +Q_{2} ({\bf k}) \right) \; .
\end{equation}
The directional dependence is encoded in the functions $Q_{1} ({\bf k})$ and $Q_{2} ({\bf k})$, 
which originate from the corrections to the linear and next-to-linear orders in the parameter $h$, respectively, 
\begin{equation} \label{LR}
Q_{1} ({\bf k})= -\frac{\pi}{k} \hat{k}_i \hat{k}_j \left (\partial_i \partial_j \eta_{\star} -\frac{1}{3} 
\delta_{ij} \partial_k \partial_k \eta_{\star} \right) \; , 
\end{equation} 
\begin{equation} 
\label{subleading}
Q_{2} ({\bf k})=-\frac{3}{2} (\hat{{\bf k}} {\bf v})^2 \; ,
\end{equation} 
where $\hat{\bf k} = {\bf k}/k$.
First, let us consider the leading order contribution $Q_{1} ({\bf k})$. We expand it 
in spherical harmonics, 
\begin{equation} \label{harmonics}
Q_{1} ({\bf k}) =a(k)\sum_{LM} q_{LM} Y_{LM} (\hat{{\bf k}}) \; ,
\end{equation}
where 
\begin{equation}
\label{feb15-10}
a(k) = k^{-1} \; .
\end{equation}
By comparing \eqref{LR} with \eqref{harmonics}, one concludes that the anisotropic coefficients $q_{LM}$ are 
Gaussian variables, 
since they are linearly related to the derivatives of the end-of-roll time $\eta_{\star} ({\bf x})$, 
which is the Gaussian field. 
We keep very long modes of  $\delta \eta_* ({\bf x})$ with $p < H_0$, where $H_0$ is the present value of the
Hubble parameter.  At shorter wavelengths the field $\delta \eta_{\star} ({\bf x})$ 
gets averaged out. Therefore, the expression in parenthesis in \eqref{LR} should be treated as a constant
tensor throughout our part of the Universe; retaining its dependence on ${\bf x}$ would result in
effects suppressed by $H_0/k$. For this reason,
only the quadrupole of the general type survives in Eq.~\eqref{harmonics}. Neither its direction nor precise
magnitude can be predicted because of the cosmic variance. Yet its variance
in the ensemble of Universes like ours is calculable. One makes use of Eq.~\eqref{feb15-3} and finds
\begin{equation} \label{stat}
\langle q_{2M} q^{\star}_{2M'} \rangle =\frac{\pi h^2 H^2_0}{25} \delta_{MM'} \; . 
\end{equation} 
For  similar reason, the second contribution $Q_2 ({\bf k})$
also represents the quadrupole statistical anisotropy, but of 
the special type. It can be expanded  in the same fashion as in~\eqref{harmonics}. As compared to the 
previous case, the amplitude $a(k)$ is independent of the wavenumber $k$. 
We will see that this fact is crucial  from the viewpoint of the CMB 
observations.
The other important 
distinction is that the quantities $q_{2M}$ are not Gaussian now. 
Therefore, it will be  convenient to work with the components 
of the ``velocity'' ${\bf v}$, which are Gaussian variables with zero means and variances 
\begin{equation} 
\label{variancev2}
\langle v^2_{i} \rangle =\frac{3h^2}{8\pi^2} \ln \frac{H_{0}}{\Lambda} \; .
\end{equation} 
Here the present value of the Hubble parameter and the constant $\Lambda$ appear as the ultraviolet 
and infrared cutoffs, respectively. 
The quantities $q_{2M}$ are then given by 
\begin{equation} 
\label{subleadingq}
q_{2M} =-\frac{4\pi v^2}{5} Y^{\star}_{2M} (\hat{{\bf v}}) \; ,
\end{equation}
where $\hat{{\bf v}}={\bf v}/v$ is the unit vector in the direction of the ``velocity'' ${\bf v}$.

Equations~\eqref{feb15-10},~\eqref{stat},~\eqref{variancev2} and~\eqref{subleadingq} are the starting point of our analysis of
the statistical anisotropy in the CMB within the sub-scenario A. 

\subsubsection{Sub-scenario B} 
\label{sec:scenarioB}

Let us now turn to the second sub-scenario, in which the phase perturbations are
still subhorizon at the end of the conformal rolling stage and exit the cosmological horizon
later~\cite{Ramazanov}.
This case is rather non-trivial. First, we need to make certain assumptions about the intermediate stage following 
conformal rolling. Barring fine-tuning, we consider the intermediate stage very long as compared to the 
interesting cosmological scales, namely,
\begin{equation}
r \equiv k(\eta_{1} -\eta_{\star}) \gg 1 \; , 
\label{feb15-11}
\end{equation}
where $\eta_1$ is the end-time of the intermediate stage, at which the modes $\delta \theta$ exit
the cosmological horizon. In order not to  
modify the flat power spectrum generated by the 
end of conformal rolling, one assumes that the dynamics of the phase perturbations $\delta \theta$ is nearly 
Minkowskian at the intermediate stage,
so that they obey the field equation
\begin{equation}\label{intermediate}
\partial^2_{\eta}\delta \theta -\partial_i \partial_i \delta \theta =0 \; .
\end{equation}
The latter condition appears very restrictive. However, there are at least two cosmological scenarios where 
it is obeyed. One is the bouncing Universe with the super-stiff equation of state, 
$p \gg \rho$~\cite{smooth,newekpy,lehners}; interestingly, cosmological
contraction with stiff equation of state is inherent also in the pseudo-conformal Universe model of Ref.~\cite{Hinterbichler}.
Another is Genesis~\cite{Trincherini}, where the Universe stays static for a very long period of time before 
it starts to expand rapidly. 

Under the above assumptions, the solution to Eq.~\eqref{intermediate} supplemented with the initial 
conditions \eqref{phaseperturb} and \eqref{freezeout} is given by~\cite{Ramazanov} 
\begin{equation}
\delta \theta ({\bf x}, \eta_1)=\frac{h}{4\pi^{3/2}} \int \frac{d^3 k}{\sqrt{k}} e^{i{\bf kx}} A_{\bf k} I +h.c. \; ,
\end{equation}
where $I$ is the sum of two incoherent waves coming from the direction $\hat{{\bf k}}$ and from the opposite 
direction, 
\begin{equation}
I=\frac{1}{2k} \left( e^{i\psi_{+}} \left[1-\hat{{\bf k}} {\bf v}^{(+\hat{{\bf k}})}+r (\delta_{ij} -\hat{k}_i \hat{k}_j)\partial_i 
v^{(+\hat{{\bf k}})}_j \right] +e^{i\psi_{-}} \left(1-\hat{{\bf k}} {\bf v}^{(-\hat{{\bf k}})} \right) \right) \; .
\label{feb15-13}
\end{equation}
Here 
\begin{equation}
\nonumber
\psi_{+} =\psi_{+} ({\bf x}, \hat{{\bf k}})=k\eta_1 -2k\eta_{\star} 
({\bf x} +\hat{{\bf k}}r), \quad \psi_{-}=-k\eta_1 \; ,
\end{equation} 
upper labels $(+\hat{{\bf k}})$ and $(-\hat{{\bf k}})$ denote quanities calculated 
at the points ${\bf x} +\hat{\bf k}r$ and ${\bf x}-\hat{\bf k}r$, respectively.
Note that Eq.~\eqref{feb15-13} is valid at the horizon exit, so it determines, in fact, the properties
of the adiabatic perturbations (again modulo possible additional non-Gaussianity). Note also that
under the assumption \eqref{feb15-11}, the modes of $\delta \eta_*$ relevant in
Eq.~\eqref{feb15-13} are indeed longer than the size of the visible Universe.

Proceeding as in Section~\ref{sec:scenarioA}, we arrive at the power spectrum~\eqref{feb15-20}, but now with
\begin{equation} \label{LRR}
Q ({\bf k})= \hat{k}_i \left (v^{(+\hat{{\bf k}})}_i-v^{(-\hat{{\bf k}})}_i \right) \; .
\end{equation}
In this case all coefficients $q_{LM}$ are generically 
different from zero. Their variances in an ensemble of Universes
are calculated by making use of Eq.~\eqref{feb15-3}
and are given by~\cite{Ramazanov} 
\begin{equation} \label{statcoef}
 \langle q_{LM} q^{\star}_{L'M'} \rangle =\frac{3h^2}{\pi} \frac{1}{(L-1)(L+2)} \delta_{LL'} \delta_{MM'} \; .
\end{equation} 
Notably, the function $a(k)$ does not depend on the momentum, i.e. 
\begin{equation}
a(k) = 1 \; .
\label{feb15-22}
\end{equation}
Equations \eqref{statcoef} and \eqref{feb15-22} determine the statistical anisotropy 
of the primordial adiabatic perturbations in the
sub-scenario B in terms of a single unknown parameter $h^2$.

\section{Estimators} 
\label{sec:estimator}
\subsection{Model-independent analysis} 
\label{sec:modelindependent}

Let us first apply the quadratic maximum likelihood (QML) method to construct the estimators for the coefficients 
$q_{ML}$. Here we define the latter in a model-independent way, assuming only that the dependence on the wavenumber 
$k$ can be factorized as in Eq.~\eqref{power}. We closely follow the technique 
developed in Ref.~\cite{Duncan}. 
In Section~\ref{sec:parameter}
we use the same ideas when constructing the estimator 
for the parameter $h^2$. 

In what follows we use the harmonic representation 
for the temperature fluctuations and their covariances unless the opposite stated. The log-likelihood of the observed 
CMB temperature map $\hat{{\bf \Theta}}$ is given by
\begin{equation}\label{loglike}
-{\cal L} (\hat{{\bf \Theta}}| {\bf q})=\frac{1}{2} \hat{{\bf \Theta}}^{\dagger} {\bf C}^{-1} 
\hat{{\bf \Theta}} +\frac{1}{2} \ln \det {\bf C} \; , 
\end{equation} 
where ${\bf q}$ is the vector of coefficients $q_{LM}$; the covariance matrix ${\bf C}$ incorporates the theoretical covariance 
corresponding to the signal as 
well as the instrumental 
noise, ${\bf C}={\bf S}+{\bf N}$. The theoretical covariance is given by 
\begin{equation} 
\nonumber
S_{lm;l'm'}=\langle \Theta_{lm} {\Theta}^{\star}_{l'm'} \rangle = 
i^{l-l'}\frac{2}{\pi} \int d{\bf k} \Delta_{l} (k) \Delta_{l'} (k) 
P_{\zeta} ({\bf k}) Y^{\star}_{lm} (\hat{{\bf k}}) Y_{l'm'} (\hat{{\bf k}}) \; .
\end{equation}
Here $\Theta_{lm}$ are the theoretical temperature fluctuations of the CMB sky 
$\delta T({\bf n})$ in the harmonic representation,
\begin{equation}
\nonumber
\Theta_{lm}= \int d \Omega  \delta T ({\bf n}) Y^{\star}_{lm} ({\bf n}) \; ,
\end{equation} 
and $P_{\zeta} ({\bf k})$ is the power spectrum of the primordial perturbations; $\Delta_l (k)$ are transfer functions.
Under the assumption of the statistical anisotropy, we write the theoretical covariance as follows, 
\begin{equation}
\nonumber 
{\bf S}={\bf S}^i +\delta {\bf S} \; , 
\end{equation}
where the leading contribution ${\bf S}^i$ comes from the isotropic signal well fitted by the $\Lambda$CDM model; the 
effects of the statistical isotropy violation are incorporated into $\delta {\bf S}$. 
The matrix ${\bf S}^i$ is diagonal in the harmonic representation, 
\begin{equation} \label{standard}
S^{i}_{lm;l'm'}=C_l \delta_{ll'} \delta_{mm'} \; .
\end{equation}
where $C_l$ is the standard CMB angular spectrum. The matrix $\delta {\bf S}$ is given by
\begin{equation}\label{covan}
\delta S_{lm;l'm'}=i^{l'-l}C_{ll'}\sum_{LM} q_{LM} \int d \Omega_{\bf k} Y^{\star}_{lm} (\hat{{\bf k}}) 
Y_{l'm'} (\hat{{\bf k}}) Y_{LM}(\hat{{\bf k}}) \; ,
\end{equation}
where 
\begin{equation}\label{transfer}
C_{ll'}=4 \pi \int d \ln k \Delta_{l} (k) \Delta_{l'} (k) a(k) {\cal P}_{\zeta} (k) \; .
\end{equation}
The integral of three spherical harmonics reads
\begin{equation} \label{Klebsh}
\int d \Omega_{\bf k} Y^{\star}_{lm} (\hat{{\bf k}}) 
Y_{l'm'} (\hat{{\bf k}}) Y_{LM}(\hat{{\bf k}})=(-1)^{m'} G^{L}_{ll'}C^{LM}_{lm;l'-m'} \; ,
\end{equation}
where $C^{LM}_{lm;l',-m'}$ are the Clebsch-Gordan coefficients and 
\begin{equation} 
\nonumber
G^{L}_{ll'} \equiv \sqrt{\frac{(2l+1)(2l'+1)}{4\pi (2L+1)}}C^{L0}_{l0l'0} \; .
\end{equation}

Normally, the estimators for the coefficients $q_{LM}$ are 
determined by equating the derivative of the log-likelihood to zero, 
\begin{equation}
\nonumber 
\frac{\partial {\cal L}}{\partial {\bf q}^{\dagger}}=0 \; .
\end{equation}
However, the covariance matrix ${\bf C}$ is not sparse 
and direct calculations are 
too costly. Thus, we need an appropriate approximation to work with. 
At this point we make use of
the QML approach.
Assuming that the statistical anisotropy is weak, we expand the log-likelihood derivative to the linear order 
in ${\bf q}$,
 \begin{equation}\label{linexp}
\frac{\partial  {\cal L}}{\partial {\bf q}^{\dagger}}=\left.\frac{\partial {\cal L}}{\partial {\bf q}^{\dagger}} \right|_{0}
+\left.\frac{\partial^2 {\cal L}}{\partial {\bf q}^{\dagger}\partial {\bf q}} \right|_{0} {\bf q}\; .
\end{equation}
We replace the second derivative of the log-likelihood in this expansion
by its expectation value~\cite{Duncan}, 
\begin{equation} \label{Fisher}
\left\langle \frac{\partial^2 {\cal L}}{\partial {\bf q} \partial {\bf q}^{\dagger}} \right\rangle =
-\left\langle \frac{\partial {\cal L}}{\partial {\bf q}}\frac{\partial {\cal L}}{\partial {\bf q}^{\dagger}} \right\rangle=
-{\bf F} \; , 
\end{equation} 
where ${\bf F}$ is the Fisher matrix. The first equality in Eq.~\eqref{Fisher} follows from the normalization condition
\begin{equation}
\nonumber 
\int \exp \left ({\cal L} \right)  d \hat{{\bf \Theta}}=1 \; .
\end{equation} 
In what follows we use the derivatives of the log-likelihood calculated under the assumption of the 
statistical isotropy unless the opposite stated, and omit the subscript ``0''.
The first derivative of the log-likelihood is then given by
\begin{equation}
\nonumber
\frac{\partial {\cal L}}{\partial {\bf q}^{\dagger}}=
\frac{1}{2} \hat{{\bf \Theta}}^{\dagger} \left( {\bf C}^i \right)^{-1} 
\frac{\partial {\bf C}}{\partial {\bf q}^{\dagger}} \left({\bf C}^i \right)^{-1} \hat{{\bf \Theta}} 
-\frac{1}{2} \mbox{Tr} \left ( \left( {\bf C}^i \right)^{-1} \frac{\partial {\bf C}}{\partial {\bf q}^{\dagger}} 
\right ) \; ,
\end{equation} 
where ${\bf C}^{i}$ is the statistically isotropic covariance incorporating the noise, ${\bf C}^{i}={\bf S}^{i} +{\bf N}$. 
The second term in the right hand side of this equation is, in fact,
the average of the first term over the realizations of CMB. This follows from the identity 
$\mbox{Tr} {\bf A}= \langle \bar{{\bf x}}^{\dagger} {\bf A} {\bf C}^{-1} \bar{{\bf x}} \rangle$,
where ${\bf A}$ is any matrix and ${\bf x}$ is a vector of Gaussian random variables with the covariance ${\bf C}$. 
Thus, one writes
\begin{equation} \label{like}
\frac{\partial {\cal L}}{\partial {\bf q}^{\dagger}}={\bf h}-\langle {\bf h} \rangle \; ,
\end{equation}
where
\begin{equation}\label{quadratic}
{\bf h}=\frac{1}{2} \bar{{\bf \Theta}}^{\dagger} 
\frac{\partial {\bf C}}{\partial {\bf q}^{\dagger}} 
\bar{{\bf \Theta}} \; ,
\end{equation}
and the quantities $\bar{{\bf \Theta}}$ are the inverse-variance filtered CMB harmonics calculated in the absence 
of the statistical anisotropy,
\begin{equation}\label{invvar}
\bar{{\bf \Theta}}=\left( {\bf S}^i +{\bf N} \right)^{-1} \hat{{\bf \Theta}} \; .
\end{equation}
By substituting Eqs.~\eqref{Fisher} and \eqref{like}
into Eq.~\eqref{linexp} and equating the result to zero, we obtain the QML estimator, 
\begin{equation} \label{HLestimator}
{\bf q}=({\bf  F})^{-1} ({\bf h} -\langle {\bf h} \rangle ) \; . 
\end{equation}
In what follows we use the Fisher matrix calculated in the full sky and homogeneous noise 
approximation~\cite{Duncan}. It has only diagonal elements, which do not depend on $M$, 
\begin{equation} \label{fishapprox}
F_{LM;L'M'} =\delta_{LL'} \delta_{MM'} w\sum_{l, l'} 
\frac{(2l+1)(2l'+1)}{8\pi}\left (
\begin{array}{ccc} 
L & l & l'\\
0 & 0 & 0
\end{array} 
\right )^2 \frac{C^2_{l l'}}{C^{tot}_{l} C^{tot}_{l'}} \; ,
\end{equation}  
where $C^{tot}_l$ is the sum of the standard CMB spectrum $C_l$ and the noise spectrum $N_l$. 
The constant $w$ denotes the uncut fraction of the sky. We include this factor to achieve better agreement between the approximate
Fisher matrix and the exact one defined as the average over the ensemble of simulated maps with the real sky coverage 
and inhomogeneous noise.

\subsection{Estimator for the parameter $h^2$} 
\label{sec:parameter}

In the framework of the conformal rolling scenario, the coefficients $q_{LM}$ are random variables with zero expectation values.
To the linear order in  $h$ they are Gaussian and have 
variances given by Eq.~\eqref{stat} or \eqref{statcoef}. The 
variances depend on the constant $h^2$, which is the only parameter of the model. This makes it possible to constrain the
conformal rolling scenario 
from the non-observation of the cosmological statistical anisotropy. We again use the QML method to construct 
the estimator for the parameter $h^2$. We do that starting from the Gaussian hypothesis about the coefficients $q_{LM}$. 
This hypothesis is particularly appropriate 
in the context of the sub-scenario B. The non-Gaussian $q_{2M}$'s appear 
in the sub-scenario A to the subleading order; we comment on this case in the end of this Section.

The probability density of the coefficients $q_{LM}$ for a given value 
of $h^2$ is
\begin{equation} 
\nonumber
{\cal W} ({\bf q}|h^2) \sim \frac{1}{\sqrt{\det {\bf Q}}} \exp \left(-\frac{1}{2} {\bf q}^{\dagger} {\bf Q}^{-1} {\bf q} \right) \; .
\end{equation} 
Here the matrix ${\bf Q}$ is the covariance of the anisotropy parameters, $Q_{LM;L'M'} \equiv \langle q_{LM} q^{\star}_{L'M'} \rangle$.
To obtain the likelihood of the observed CMB with respect to the parameter $h^2$, one integrates 
the product of two probability densities over the set of the parameters $q_{LM}$,
\begin{equation} \label{totprobdens}
{\cal W} (\hat{{\bf \Theta}} | h^2)=\int {\cal W} (\hat{{\bf \Theta}}| {\bf q}){\cal W} ({\bf q}|h^2) d{\bf q} \; ,
\end{equation} 
where ${\cal W} (\hat{{\bf \Theta}}, {\bf q}) = \mbox{exp} [{\cal L}  (\hat{{\bf \Theta}}, {\bf q})]$ 
and ${\cal L}$ is the log-likelihood introduced in Eq.~\eqref{loglike}.
Following the main idea of the QML estimation, we expand the log-likelihood to the second order in ${\bf q}$,
\begin{equation}
\nonumber
{\cal L}={\cal L}_0 +\frac{\partial {\cal L}}{\partial {\bf q}} {\bf q}-\frac{1}{2} {\bf q}^{\dagger}
{\bf F} {\bf q} \; ,
\end{equation} 
where we again replaced the second derivative by its expectation value over the CMB isotropic realizations. 
Now the integral in Eq.~\eqref{totprobdens} 
takes a simple Gaussian form and can 
be straightforwardly evaluated, 
\begin{equation} \label{finaldensprob}
{\cal W} \sim \frac{1}{\sqrt{\det ({\bf FQ} +{\bf I})}} 
\exp \left(\frac{1}{2} \frac{\partial {\cal L}}{\partial {\bf q}^{\dagger}} ({\bf  F}+{\bf Q}^{-1})^{-1}
\frac{\partial {\cal L}}{\partial {\bf q}} \right) \; .
\end{equation}
Maximizing \eqref{finaldensprob} with respect to the parameter $h^2$,
\begin{equation} 
\nonumber
 \frac{\partial \ln {\cal W} (\hat{{\bf \Theta}}|h^2)}{\partial h^2}=0 \; ,
\end{equation}
we obtain the equation for the estimator of $h^2$,
\begin{equation} 
\nonumber
 \mbox{Tr} \left(({\bf FQ}+{\bf I})^{-1}{\bf F} \frac{\partial {\bf Q}}{\partial h^2} \right)= 
 \frac{\partial {\cal L}}{\partial {\bf q}^{\dagger}} ({\bf FQ} +{\bf I})^{-1} \frac{\partial {\bf Q}}{\partial h^2} 
({\bf FQ}+{\bf I})^{-1} \frac{\partial {\cal L}}{\partial {\bf q}}\; .
\end{equation} 
In the full sky and homogeneous noise approximation, the Fisher matrix~\eqref{fishapprox} 
is diagonal,
\begin{equation} 
\nonumber
F_{LM;L'M'}=F_{L} \delta_{LL'} \delta_{MM'} \; .
\end{equation}
The matrix $Q$ has the same property, 
\begin{equation}
\nonumber
Q_{LM;L'M'}= \tilde{Q}_L h^2 \delta_{LL'} \delta_{MM'} \; ,
\end{equation}
where we introduce the quantities $\tilde{Q}_{L}$ which do not depend on the parameter $h^2$.
Then the equation determining the estimator takes the form
\begin{equation} \label{estim}
h^2 \sum_{L} \frac{(2L+1)F^2_L \tilde{Q}^2_L}{(1+F_L \tilde{Q}_L h^2)^2}=
\sum_L \frac{(2L+1) F_L \tilde{Q}_L}{(1+F_L \tilde{Q}_L h^2)^2}
(F_L C^{q}_L -1) \; ,
\end{equation}
where we use the same notation $h^2$ for the estimator as for the parameter of the model. 
The quantities $C^{q}_{L}$ entering Eq.~\eqref{estim} are given by 
\begin{equation}
\label{feb17-1}
C^{q}_{L} =\frac{1}{2L+1} \sum_{M} |q_{LM} |^2 \; , 
\end{equation}
where the coefficients $q_{LM}$ are defined by Eq.~\eqref{HLestimator}. 

Note that it follows from Eq.~\eqref{estim}
that if the predicted statistical anisotropy is of the quadrupole form,
i.e., with non-zero $q_{2M}$'s only, 
then
the parameter $h^2$ can be estimated simply as $h^2=C^{q}_{2}$, modulo obvious
additive and multiplicative constants. This is also clear on 
general grounds. Indeed, the rotational invariance 
requires that the estimator should be some function of $C^{q}_{2}$, i.e. $h^2=f(C^{q}_2)$. In the 
small statistical anisotropy approximation, we keep only linear terms in the Taylor expansion 
of the function $f(C^{q}_2)$. This immediately implies the quoted relationship
between $h^2$ and  $C^{q}_{2}$. 
Since no  assumptions 
about the properties of the random quantities $q_{2M}$ have been used in the latter
argument, 
it holds for non-Gaussian $q_{2M}$'s, which describe the quadrupole 
of the 
special type, see Eq.~\eqref{subleadingq}. The only qualification is that 
the statistical anisotropy is of the order $h^2 \ln \frac{H_{0}}{\Lambda}$ 
in that case. Hence, 
the corresponding estimator 
reads $h^4 \ln^2 \frac{H_0}{\Lambda}=C^{q}_2$, up to multiplicative and additive
constants.


\section{Implementation and results} 
\label{sec:results}

\begin{figure}[tb!]
\begin{center}
\includegraphics[width=0.34\columnwidth,angle=-90]{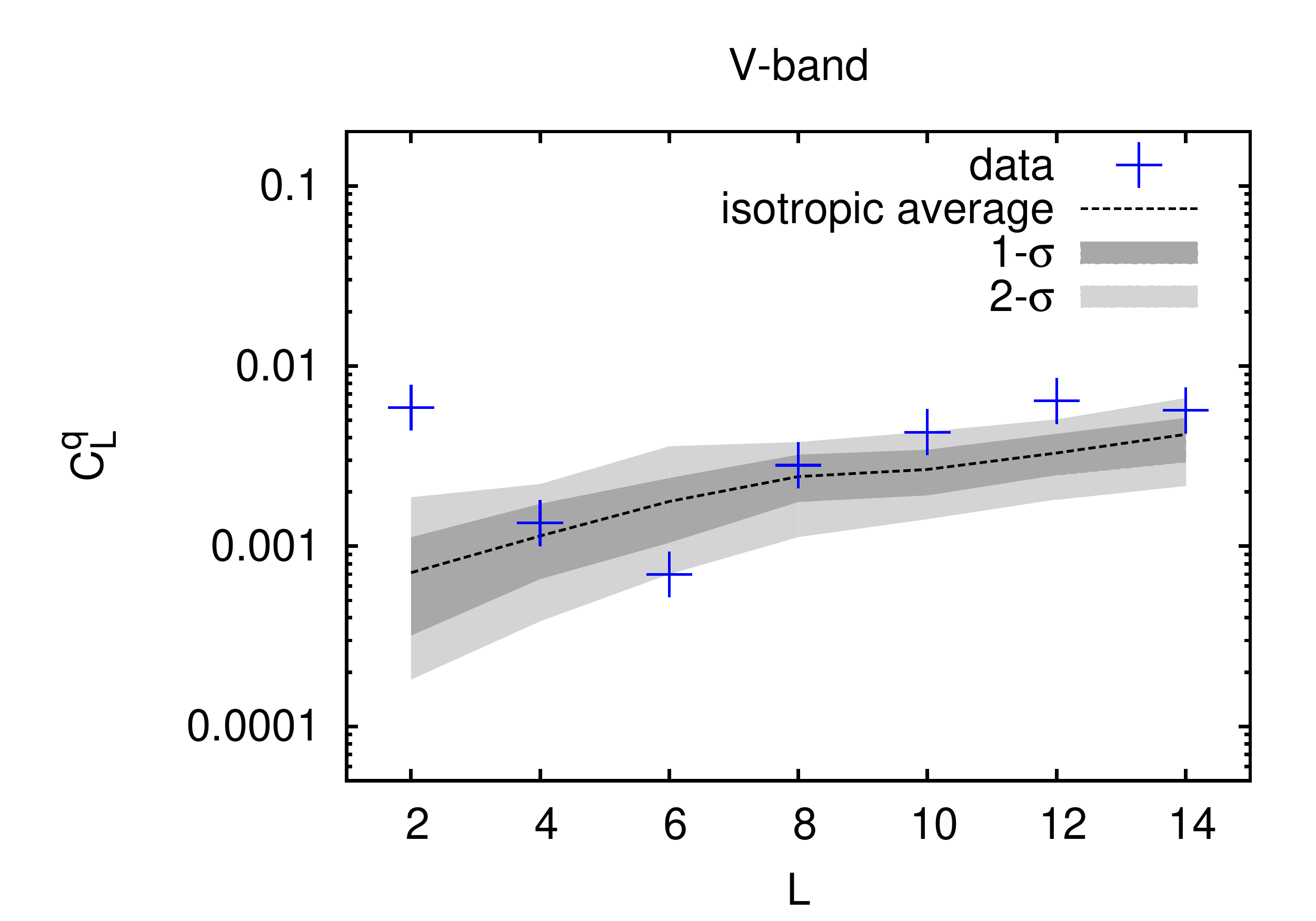}
\includegraphics[width=0.34\columnwidth,angle=-90]{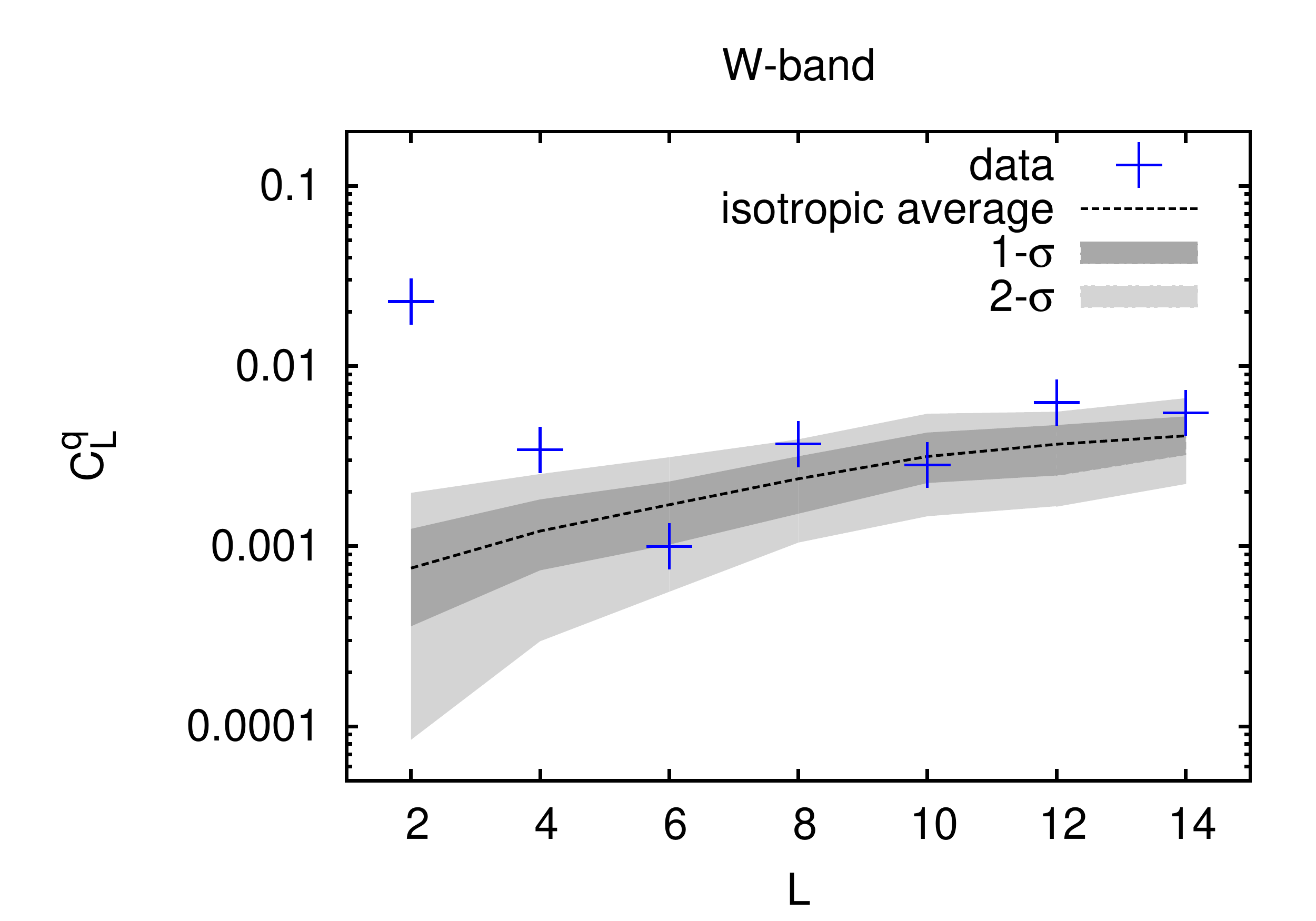}
\end{center}
\caption{$C^{q}_{L}$ of the $q_{LM}$ reconstruction for the $V$ (left) and $W$ (right) bands of the 
seven-year WMAP data. This analysis assumes $a(k)=1$ in Eq.~\eqref{power}. 
The $1\sigma$ (dark grey) and $2\sigma$ (light grey) confidence intervals are calculated using MC 
simulated statistically isotropic maps. The analysis is performed with the WMAP temperature analysis mask 
and $l_{max}=400$. \label{CL}}
\end{figure}

We search for the statistical anisotropy using WMAP
seven-year maps~\cite{Jarosik:2010iu,Komatsu:2010fb}. We study the $V$ and
$W$ band data at 61 and 94~GHz.  The first step is to implement inverse-variance filtering defined by
Eq.~\eqref{invvar}. We write that equation in the form appropriate for applying the
conjugate gradient technique,
\begin{equation} \label{supereq}
[\left({\bf S}^i \right)^{-1} + \tilde{{\bf Y}}^{\dagger} {\bf N}^{-1} \tilde{{\bf Y}}]{\bf S}^i\bar{{\bf \Theta}}=
\tilde{{\bf Y}}^{\dagger}
{\bf N}^{-1} \hat{{\bf \Theta}} \; .
\end{equation}
Here we use the pixel representation for the noise covariance ${\bf N}$ and the 
observed CMB temperature $\hat{{\bf \Theta}}$. The matrix $\tilde{{\bf Y}}$ relates the harmonic space
covariance and the observed map,
\begin{equation} 
\nonumber
\tilde{Y}_{i,lm}=B_l Y_{lm} (\vartheta_i , \varphi_i) \; ,
\end{equation}
where $B_l$ are the beam transfer functions and
$i$ labels pixels. We use the foreground reduced
$V$ and $W$ seven-year maps~\cite{nasa} provided in
HEALPix format~\cite{Healpix} with $N_{side}=512$. For the beam
transfer function we use the average of $V1$ and $V2$ functions for $V$ 
band and the average of $W1$, $W2$, $W3$ and $W4$ for $W$ band.

We consider the noise of the pixels uncorrelated and having the
variance $\sigma_0^2/n_{obs}$, where $\sigma_0$ is
$3.137$~mK and $6.549$~mK for $V$ and $W$ bands, respectively, and
$n_{obs}$ is specific to each pixel and tabulated in the maps. To
remove foreground contaminated pixels we use the WMAP temperature
analysis mask which leaves us with $w=78\%$ of the sky. We take the noise
covariance to be infinite (inverse noise is zero) for masked
pixels. The noise model ${\bf N}^{-1}$ is constructed using noise
covariance and template maps for removing monopole and dipole
contributions.

To evaluate the confidence intervals, inverse filtering should be performed
on both data and large number of simulated maps. Thus, the
system~\eqref{supereq} must be well preconditioned. Following Ref.~\cite{Duncan}, we make use of the multigrid
preconditioner, first proposed by Smith et. al. in Ref.~\cite{Smith}. It
is known to be the fastest to date and has a typical cost of ten
minutes when evaluated to $l_{max}=1000$.
 
Next, we compute the quantities $h_{LM}$ given by 
Eq.~\eqref{quadratic}. Using Eqs.~\eqref{covan} and~\eqref{Klebsh}, we write them as follows,
\begin{equation} \label{hLM}
h_{LM} =\frac{1}{2} \sum_{lm;l'm'} (-1)^{m'} i^{l'-l} \bar{\Theta}^{\star}_{lm} \bar{\Theta}_{l'm'} C_{ll'} G^{L}_{ll'}C^{LM}_{lm;l'-m'} \; .
\end{equation}
We calculate the Clebsch-Gordan coefficients using the GSL~\cite{gsl}
and Slatec~\cite{slatec} libraries. The summation in~\eqref{fishapprox} and~\eqref{hLM} is performed up to $l_{max}=400$. 
We use the publicly available Boltzman code (CAMB)~\cite{Lewis} to compute the quantities $C_{ll'}$.

\begin{figure}[tb!]
\begin{center}
\includegraphics[width=0.34\columnwidth,angle=-90]{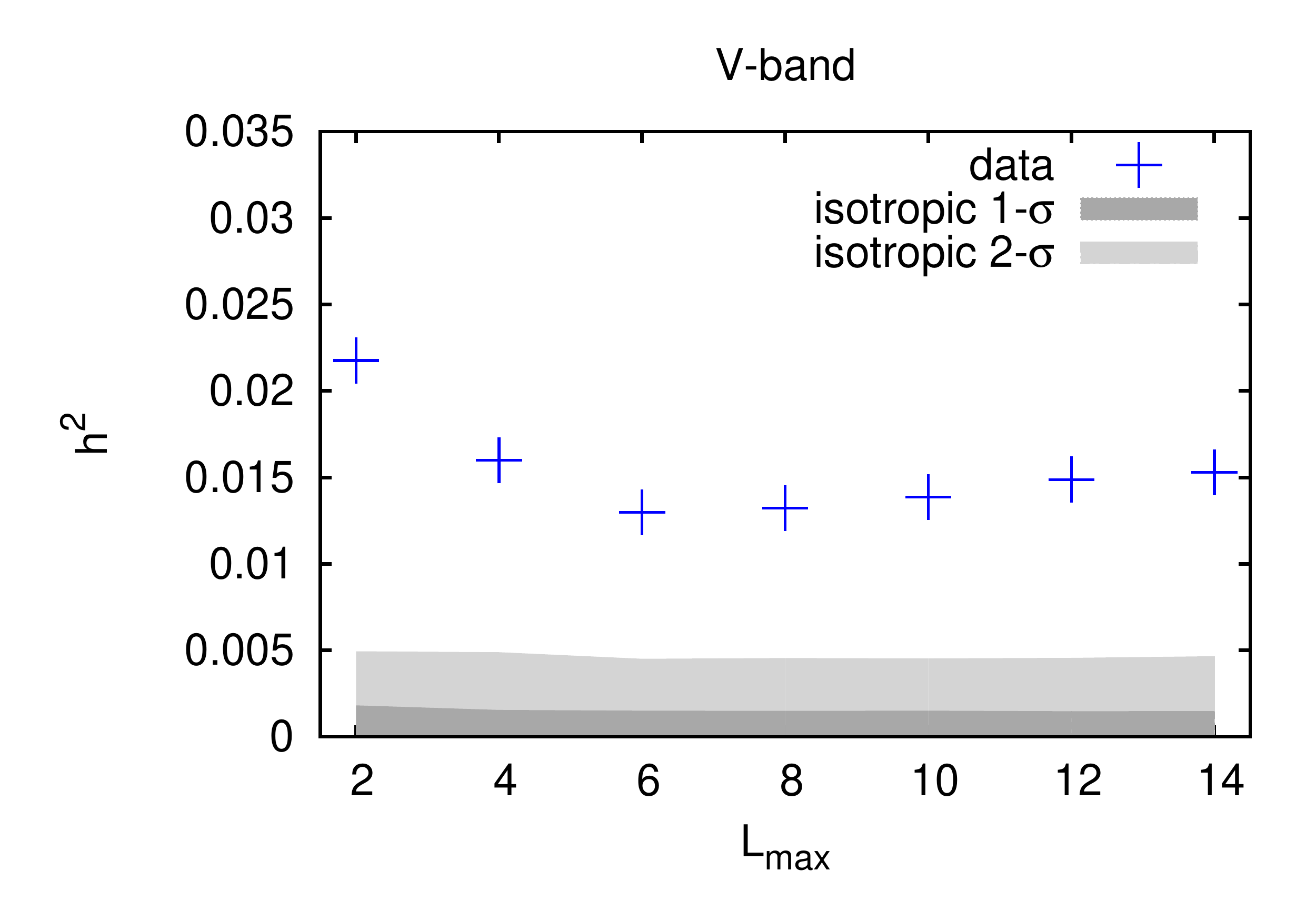}
\includegraphics[width=0.34\columnwidth,angle=-90]{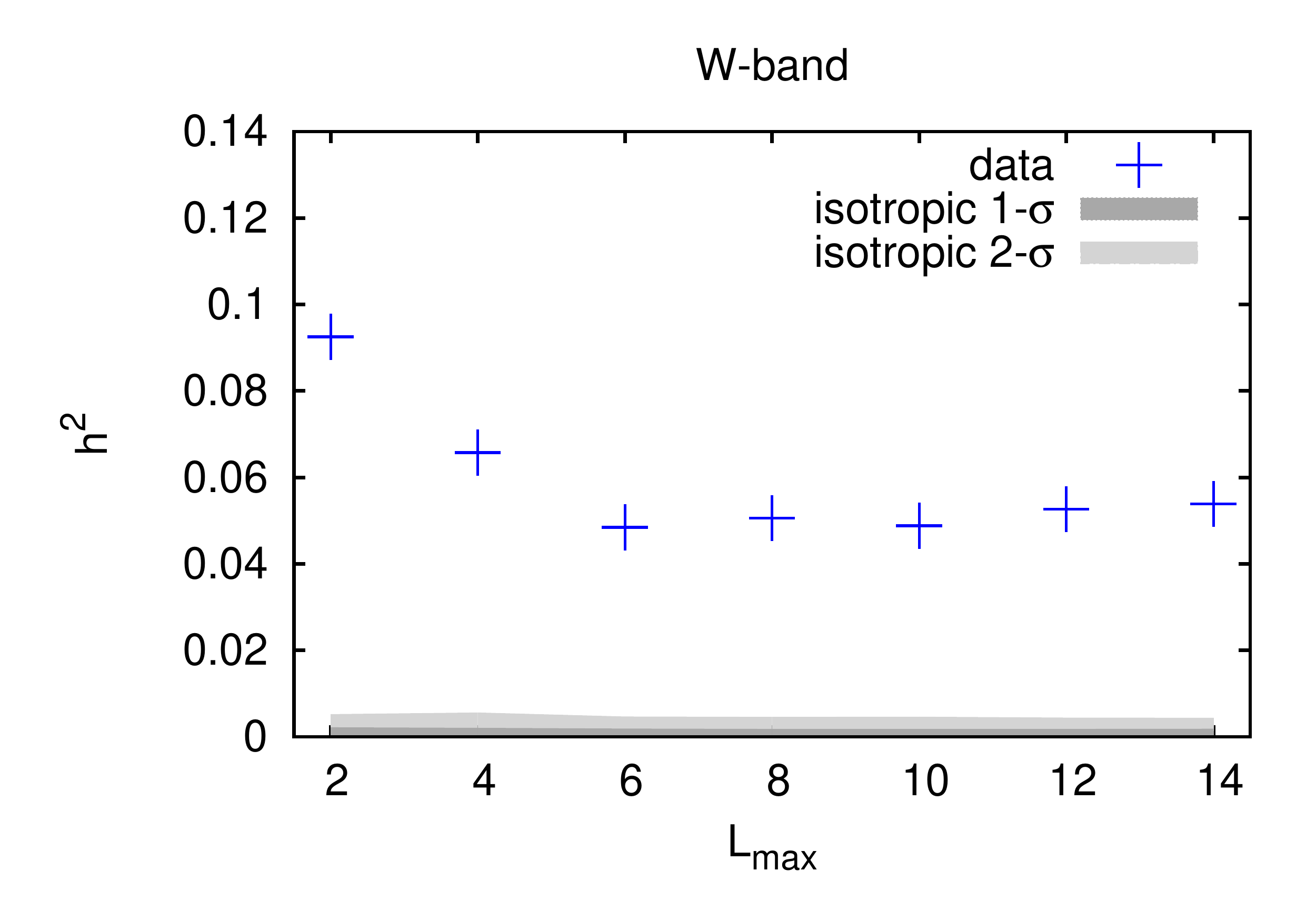}
\end{center}
\caption{Parameter $h^2$ of the sub-scenario B reconstructed from the WMAP $V$ band (left) and $W$ band
(right). The $1\sigma$ (dark grey) and $2\sigma$ (light
grey) confidence intervals obtained from MC simulations are
also shown. \label{h2}}
\end{figure}

We have simulated large number of statistically isotropic Monte-Carlo~(MC) realizations of
the field $\hat{\bf \Theta}$ using WMAP noise covariance and beam
transfer functions. We store the MC maps in the same format as the
original map, and the analysis procedure explained in this Section is
applied to both data and MC maps on equal footing.

Now we can check the consistency of the observed CMB with the
hypothesis of the statistical isotropy. 
We begin with the model-independent analysis, as outlined in Section~\ref{sec:modelindependent}.
We reconstruct coefficients  $C^{q}_L$, defined by Eq.~\eqref{feb17-1}, from the
seven-year WMAP data as well as from the MC maps. The results are
presented in Fig.~\ref{CL}. They are in a good agreement with
the results obtained by Hanson and Lewis~\cite{Duncan} for the
five-year maps. In particular, we confirm the result on the large quadrupole for
the $V$ and $W$ bands. As discussed in Refs.~\cite{Groeneboom}--\cite{Eriksen},
the preferred quadrupole direction lies very close to the ecliptic
poles. Another suspicious thing is the frequency dependence of the
signal. Namely, it is non-zero in the $W$ band at much higher confidence
level than in the $V$ band. This indicates a systematic effect rather
than the cosmological origin. As discussed in Ref.~\cite{Challinor}, the
account of beam asymmetries can provide a complete explanation of the
anomaly.

Let us turn to the conformal rolling scenario. First, we consider the
version of the model with the intermediate stage (sub-scenario B).  
The statistical anisotropy is determined by Eqs.~\eqref{statcoef} and \eqref{feb15-22}.
Having the set of the
coefficients $C^{q}_L$ reconstructed from the observed CMB, we 
solve Eq.~\eqref{estim} and estimate the value of $h^2$. We
perform the analysis for the multipole numbers starting from
$L_{min}=2$ and ranging up to $L_{max}=2-14$. The results are presented in Fig.~\ref{h2}.
To evaluate the statistical errors we
use about one hundred MC simulated isotropic maps. We see
that the isotropic model is ruled out at more than $3\sigma$ confidence
level even in the $V$ band. However, the large value of $h^2$ (e.g., $h^2 \approx 0.015$
at $L_{max}=14$) is due to the anomalous quadrupole anisotropy, which is argued to have non-cosmological origin.

\begin{figure}[tb!]
\begin{center}
\includegraphics[width=0.34\columnwidth,angle=-90]{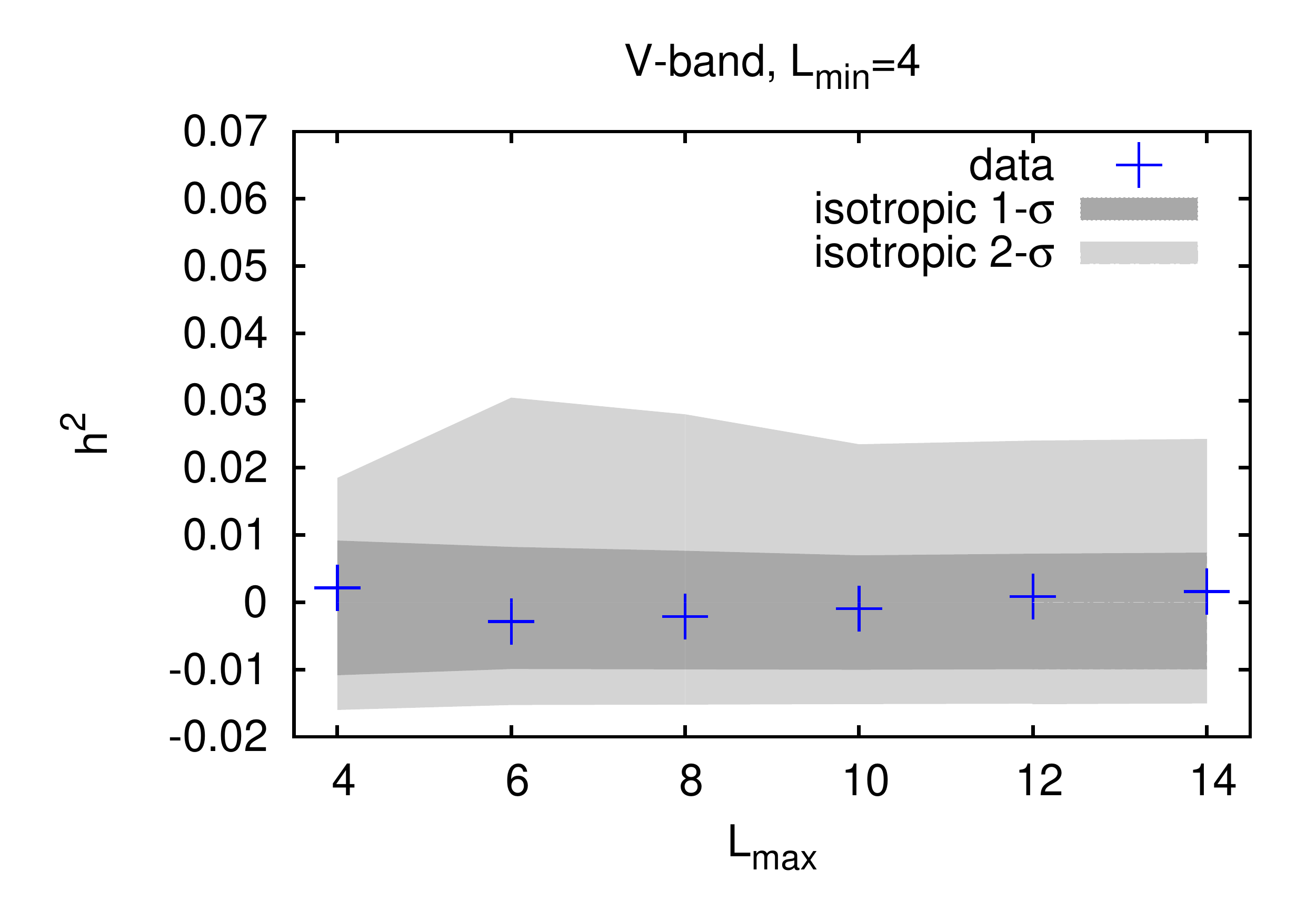}
\includegraphics[width=0.34\columnwidth,angle=-90]{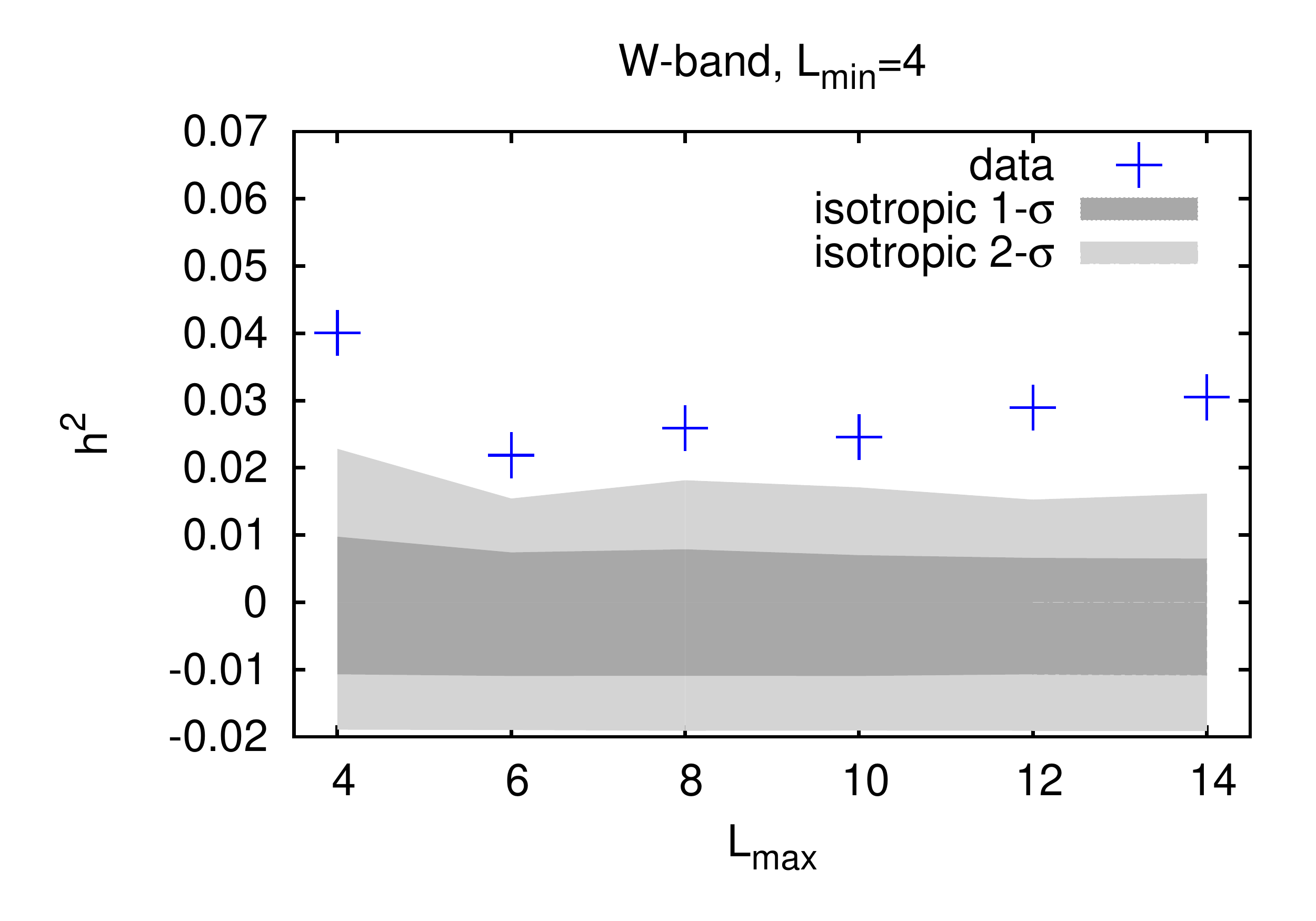}
\end{center}
\caption{Parameter $h^2$ of the sub-scenario B reconstructed from higher multipoles ($L_{min}=4$). Results are plotted
 for the WMAP $V$ band (left) and $W$ band (right). Shown are
the $1\sigma$ (dark grey) and $2\sigma$ (light grey) confidence intervals obtained from MC simulations. 
\label{h24}}
\end{figure}

\begin{figure}[tb!]
\begin{center}
\includegraphics[width=0.34\columnwidth,angle=-90]{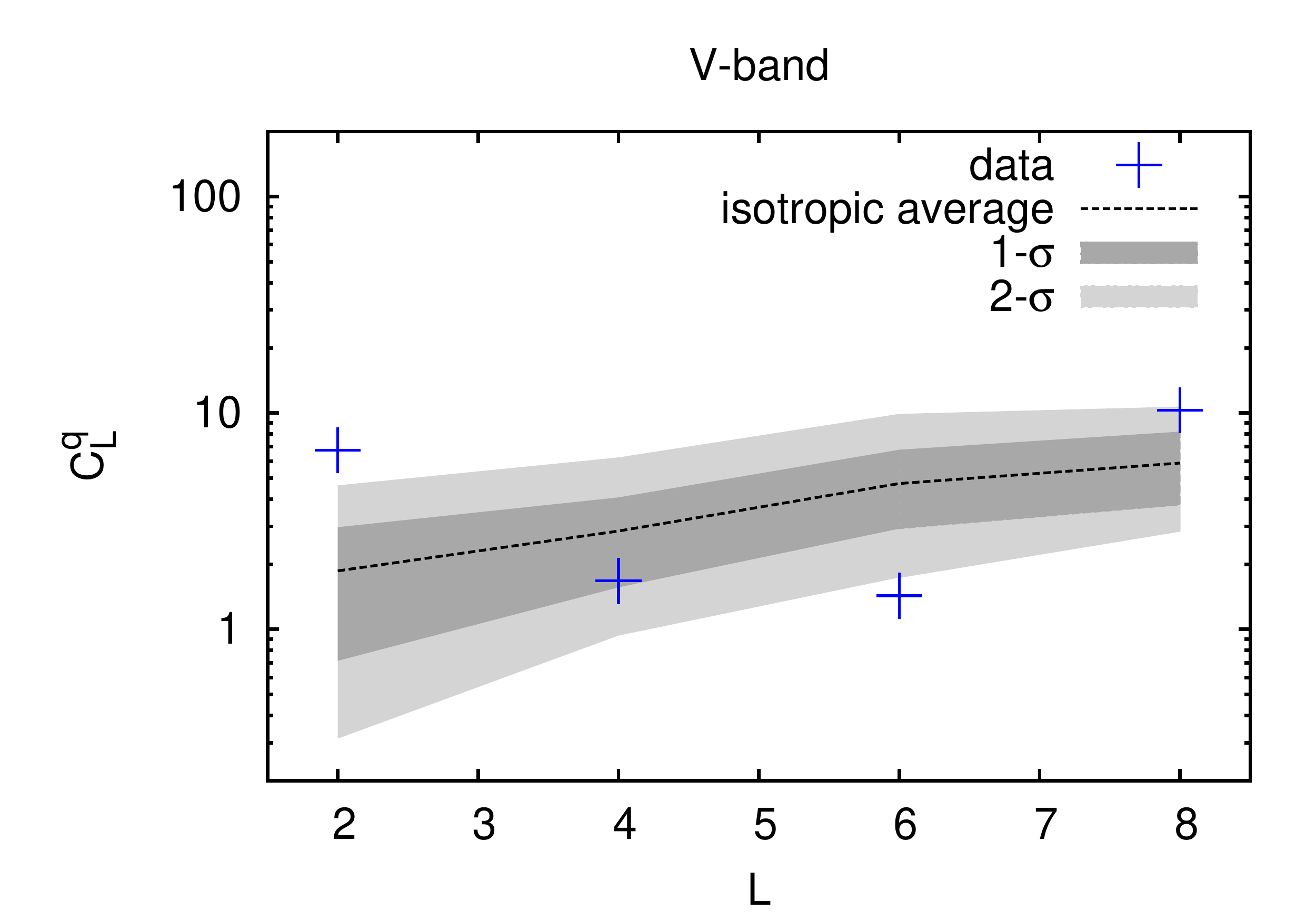}
\includegraphics[width=0.34\columnwidth,angle=-90]{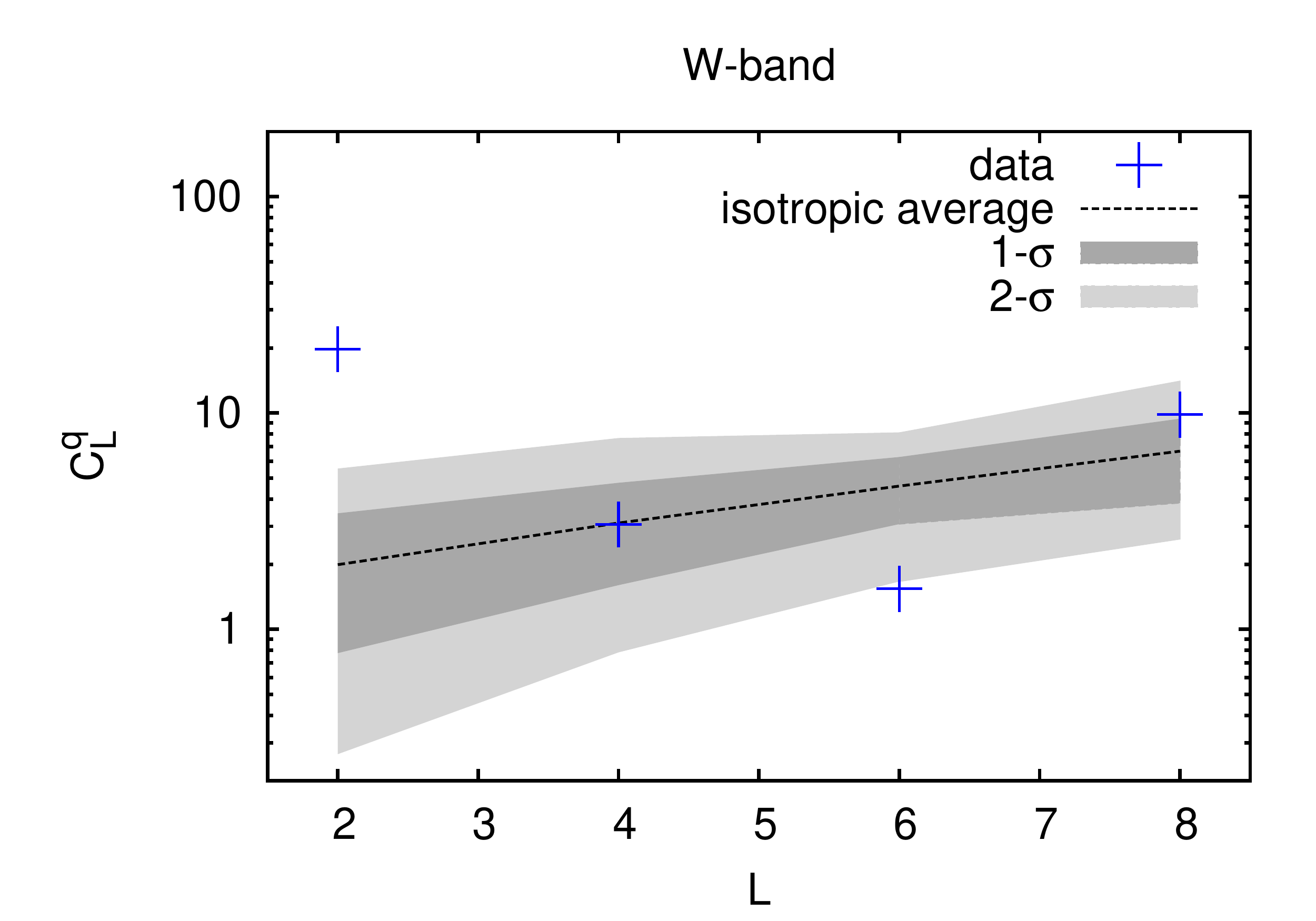}
\end{center}
\caption{
$C^{q}_{L}$ of the $q_{LM}$ reconstruction for the $V$ (left) and $W$ (right) bands of the 
seven-year WMAP. The momentum dependence of the statistical anisotropy is $a(k)=H_0 k^{-1}$.
The $1\sigma$ (dark grey) and $2\sigma$ (light grey) 
confidence intervals are calculated using MC 
simulated statistically isotropic maps. 
\label{k1}}
\end{figure}

Aiming at constraining the parameter $h^2$, we simulate a large
number of anisotropic maps for each  value of
this parameter. We adapt the approach of Ref.~\cite{Duncan}, and use the following procedure, 
adequate in the case of small statistical anisotropy.

We first simulate a seed map ${\bf \Theta}^i$ with a covariance 
${\bf S}^i$~given by Eq.~\eqref{standard}. Then we generate a set of coefficients $\{ q_{LM}
\}$ based on the value of $h^2$. The map
\begin{equation}
\nonumber
{\bf \Theta^a}=\left({\bf I}+\delta {\bf S}\left[{\bf S}^i \right]^{-1} \right)^{1/2} {\bf \Theta}^i \; , 
\end{equation}
has covariance ${\bf S}^i+\delta {\bf S}$, where $\delta {\bf
  S}$ is given by Eq.~\eqref{covan}. To the linear order in the anisotropic effects we have
\begin{equation} 
\nonumber
{\bf \Theta^a}={\bf \Theta}^i+\frac{1}{2} \delta {\bf S} [{\bf S}^i ]^{-1} {\bf \Theta}^{i} \; .
\end{equation}
Finally, we multiply the map by the beam transfer function in the
harmonic space, convert it to coordinate space and add pixel noise to get
statistically anisotropic simulated map ${\bf \hat\Theta^a}$ similar to that
observed by WMAP. To set an upper limit, we allow $h^2$ to be so large 
that for $95\%$ of simulated anisotropic maps the value of the estimated
parameter exceeds the value estimated from the observed CMB map. In this way we obtain the upper limit, which reads
\begin{equation}
\label{feb17-5}
h^2 < 0.045 \; 
\end{equation}
at the 95\% confidence level. 

In view of the likely non-cosmological origin of the anomalous quadrupole in the
statistical anisotropy of the WMAP data, one would like to constrain the parameter $h^2$
from the non-observation of higher multipoles only. One way to do that would be to
follow the same procedure as discussed in Section~\ref{sec:parameter} but keep 
the set $\{q_{2M} \}$ of the quadrupole coefficients fixed
and taken from the observational data. 
In practice, things are simpler. 
Indeed, the effects of the 
statistical anisotropy corresponding to different multipole numbers $L$, $M$ 
do not interfere with each other, at least 
in the approximation of small coefficients $q_{LM}$. To see this, we note that the theoretical reconstruction of
the Fisher matrix~\eqref{fishapprox} is diagonal. The covariances of the quantities $q_{LM}$ are diagonal as well.
As a consequence, it is straightforward to neglect the effect of the quadrupole modulation by using the estimator 
for the parameter $h^2$ as in~\eqref{estim} but with the
summation starting from $L_{min}=4$. The values of $h^2$ estimated in this way are 
plotted in Fig.~\ref{h24}. We restrict our analysis to $L_{max}=14$ and obtain that $h^2$
is consistent with zero for the V band. 
Making use of the statistical uncertainty inferred from isotropic MC maps, we obtain the upper limit 
\begin{equation}
\nonumber
h^2 < 0.040
\end{equation}
at the $95\%$ confidence level. Even though omitting the anomalous quadrupole
makes the situation cleaner (at least in the $V$ band), this constraint is similar to 
Eq.~\eqref{feb17-5}.
The reason is twofold. First,
 according to Eq.~\eqref{statcoef}, the predicted statistical anisotropy spectrum
$C^{q}_L$ decreases with $L$ as 
\[
C^{q}_L \propto \frac{2L+1}{(L-1)(L+2)} \; .
\]
Second, the errors grow with the multipole number roughly as $L$, see Fig.~\ref{CL}. 

With the Planck data available, we expect substantial improvement of the 
constraint~\eqref{feb17-5}. The reason is twofold. Hopefully, the quadrupole anomaly will be absent 
in the Planck data. Also, the range of the CMB multipoles useful in the analysis 
will be considerably extended. The error bars, whch can be 
roughly estimated by making use of the inverse Fisher matrix, 
scale with the number of multipoles 
as $l^{-2}_{max}$. This is clear from the Eq.~\eqref{fishapprox}. Taking, e.g., $l_{max}=1200$, one would be able to
reduce the error bars by about an order of magnitude. 
Hence, the non-observation of the statistical anisotropy 
will give the constraint as strong as $h^2 \lesssim 0.001$. We conclude 
that the statistical anisotropy is a promising signature from the viewpoint 
of the CMB observations in the case of the sub-scenario B.

Finally, we consider the sub-scenario A. 
To the linear order in constant $h$,
the statistical
anisotropy is of the general quadrupole type with
decreasing amplitude,  $a(k) \propto k^{-1}$. This fact is crucial for the
search for the statistical anisotropy in the CMB
sky. Indeed, the contribution to the signal $\delta {\bf S}$ is
additionally suppressed by the CMB multipole number $l$.  
This suppression is due to the fact that
the integral in Eq.~\eqref{transfer} is
saturated, roughly speaking, at $k\sim l H_0$. Effectively, it results in low statistics of the relevant CMB multipoles and 
large statistical errors, which severely restrict the opportunity to
observe the (cosmological) statistical anisotropy of the type predicted. Somewhat loosely we
apply the QML estimator to the seven-year WMAP data. 
In Fig.~\ref{k1} we show the results for $C^{q}_{L}$ of the WMAP
reconstructed coefficients $q_{LM}$, assuming $a(k)=H_0 k^{-1}$, but not restricting 
yet to the quadrupole-only $q_{LM}$. 
We apply the procedure used in the case of the sub-scenario with 
the intermediate stage, to constrain the sub-scenario A; to this end,
the quadrupole point $L=2$ in Fig.~\ref{k1} is relevant only. The limit 
on the parameter $h^2$ then reads
\begin{equation}
\nonumber 
h^2 <190
\end{equation}
at the 95\% confidence level. 
Note, however, that for large values of $h^2$, the QML procedure is
questionable. This limit can be viewed merely as an indication that the leading 
order contribution to the statistical anisotropy is in fact negligible. 
The stronger constraint comes from the subleading contribution encoded in~\eqref{subleading}. 
The reason is that the amplitude $a(k)$ is independent of the wavenumber $k$ 
in this case. Thus, the suppression at high CMB multipoles is absent, 
and the range of relevant $l$'s is extended up to $l_{max}=400$. 
Since the quantities 
$q_{2M}$ are non-Gaussian in this particular case, the constraining procedure is somewhat different. 
First, we generate the components of the 
``velocity'' ${\bf v}$ starting from a given value of the effective constant 
$h^2 \ln \frac{H_{0}}{\Lambda}$. Then, using~\eqref{subleadingq}, we calculate the 
coefficients $q_{2M}$. The quantity $C^{q}_{2}=\frac{1}{5} \sum_{M} |q_{2M}|^2$ 
constructed out of these $q_{2M}$ is compared with the one estimated 
from the seven-year WMAP data. In this way we obtain the constraint, which reads 
\begin{equation} 
h^2 \ln \frac{H_{0}}{\Lambda} <7 \;
\end{equation} 
at the 95\% confidence level. Assuming that the logarithmic enhancement is not 
particularly strong, 
we conclude that this constraint  is still weak. Note also 
that the statistical anisotropy predicted by the 
sub-scenario~A is of the same type as in some inflationary 
models~\cite{Ackerman}~--~\cite{Wagstaff}. Thus, it cannot be used to descriminate our model from 
other models of the generation of primordial perturbations. 
Fortunately, the sub-scenario~A gives rise to the 
non-Gaussianity in the trispectrum~\cite{Mironov2, Mironov1}, 
which is in the sharp contrast with the inflationary predictions. 
We leave for the future search for the corresponding signatures in the CMB sky.

\section*{Acknowledgements} 
We are indebted to V.~Rubakov  for
 numerous helpful suggestions at all stages. We are grateful to D.~ Hanson 
for kindly providing the code for inverse variance
 filtering. We would like to thank A. Lewis, P. Naselsky, M. Sazhin and 
O. Verkhodanov for fruitful discussions. The work is supported in
 part by the RFBR grants 11-02-01528a and 12-02-00653a (GR), by the
 grants of the President of the Russian Federation NS-5590.2012.2,
 MK-1632.2011.2 (GR), by the grant of the Government of 
Russian Federation 11.G34.31.0047 (GR). The numerical part of the work has been done at
 the cluster of the Theoretical Division of INR RAS.


\end{document}